\documentclass[letterpaper,times]{IONconf}

\usepackage{amsmath,amssymb}
\usepackage{graphicx}
\usepackage{url}
\usepackage{natbib}
\renewcommand{\cite}{\citep}
\usepackage[hidelinks]{hyperref}
\usepackage{enumitem}   

\title{Real-time Pre-Correlation GNSS Interference Classification with Lightweight Learned Algorithms}

\author{
    Burak~Soner$^{\dagger,\ddagger}$, Can~Aksoy$^{\ddagger}$, Erkin~Halaçlı$^{\ddagger}$%
    \vspace{1mm} \\%
    $^{\dagger}$\textit{sobu Labs, Ankara, Türkiye} \\
    $^{\ddagger}$\textit{EDGE Microwave, Istanbul, Türkiye}
}

\begin{document}

\maketitle

\section*{biography}

\biography{Burak Soner}{received his B.Sc.~in mechatronics from Sabancı University and his Ph.D.~in electrical and electronics engineering from Koç University, both in Istanbul, Türkiye. He is a multidisciplinary engineer with more than ten years of simultaneous research and professional experience in power electronics, control, wireless communication and sensing, optics, embedded systems, FPGAs, and artificial intelligence (AI) for edge devices. He recently founded his own company, \emph{sobu Labs}, in Ankara, Türkiye, providing research \& engineering services for compact DSP / AI algorithms on FPGAs and other SoCs, primarily for autonomous navigation applications, for customers in Türkiye, EU and USA.}

\biography{Can Aksoy}{received his B.Sc. and M.Sc. degrees in electronics engineering from Sabancı University, Istanbul, Türkiye, where he graduated with high honors and focused on advanced signal processing and brain–computer interface research. He has previously worked at ASELSAN and FEV Türkiye, gaining experience in embedded systems and applied signal processing. He is currently a Software Design Engineer at \emph{EDGE Microwave}, contributing to the software architecture and real-time implementation of compact CRPA systems for anti-jamming GNSS applications. His research focus is toward embedded software development, DSP algorithms, SoC-based RTOS system design, and high-reliability navigation technologies.}

\biography{Erkin Halaçlı}{received his B.Sc.~from Istanbul Technical University (İTÜ) and his M.Sc.~from Sakarya University, both in electronics engineering. He has built a career of more than 15 years at the intersection of RF engineering, antenna technologies, and secure mission-critical communication systems. He was a senior researcher at the Scientific and Technological Research Council of Türkiye (TÜBİTAK) during 2009-2016, and he later co-founded \emph{ERA RF} (2016–2024), where he led the development of high-performance RF components and antenna systems in aerospace. He is currently a founding partner of \emph{EDGE Microwave}, Istanbul, Türkiye, focusing on next-generation anti-jamming GNSS solutions such as compact CRPA systems for resilient PNT.}

\section*{Abstract}

Radio-frequency interference (RFI) remains a significant threat to GNSS in safety-critical applications. Since no single mitigation method is effective for all interferers, reliable classification is needed to select appropriate countermeasures during operation. To prevent loss of lock, RFI classification must run in real time, typically on resource-constrained embedded platforms, necessitating lightweight algorithms. While prior works realize this with simple rule-based algorithms for detecting and characterizing certain types of interferers, this approach does not scale to the broad space of all possible RFI techniques, and data-driven learned algorithms are a better fit. To satisfy these constraints, this work considers pre-correlation RFI classification with an emphasis on compact algorithms that still provide high accuracy. We first introduce a new set of lightweight input features derived from the instantaneous frequency predictions of a virtual adaptive notch filter (ANF). We observe improved classification accuracy for both narrowband and broadband non-stationary RFI by combining these new features with other spectral features from prior literature. Next, we benchmark compact learned classifiers such as gradient-boosted decision trees for accurate prediction under tight compute and memory budgets. The evaluation spans a broad set of simulated and recorded RFI events, including publicly available datasets from recent studies. Finally, we measure resource utilization for our models and for representative methods from the literature, on a compact CRPA platform (EDGE Microwave HEDGE8008).

\section{Introduction}

Global Navigation Satellite System (GNSS) receivers are increasingly exposed to radio-frequency interference (RFI) in civilian and safety-critical environments. Intentional jammers such as personal privacy devices (PPDs) and unintentional emissions from nearby RF equipment (e.g., communications, radar, switched-mode power electronics) can raise the in-band noise floor, distort correlation shapes, and ultimately cause loss of lock or degraded position, velocity, and time (PVT) performance \cite{vandermerwe_ppd_eval_2018}. Over the last decade, field campaigns and monitoring networks have documented the prevalence and diversity of such interferers, motivating practical receiver-side defenses \cite{eu_sota_rfi_2023}.

No single mitigation algorithm is uniformly effective across all interferers. Narrowband tones and slow chirps can be excised by notch filtering or transform-domain attenuation, whereas fast wideband chirps and pulsed or hopping signals tend to break the assumptions of those methods and require different countermeasures \cite{borio_benchmark_2020}. As a result, there is strong value in classification to identify the interferer type quickly and reliably. While there are post-correlation approaches that attempt to classify the type of interferer via tracking loop parameters \cite{bek2015_cairo_postcorr}, lightweight pre-correlation methods are preferable so that the receiver can select or parameterize an appropriate mitigation chain before tracking loops start to deteriorate. 

Early classification approaches have relied on hand-engineered, rule-based features derived from spectrum characteristics or periodic structures, paired with simple decision logic \cite{kim2012gnss, borio_2014_magazine}. While effective for well-separated classes and on high-SNR scenarios, these methods do not scale well as interferer diversity grows and as operating conditions shift (front-end bandwidth, automatic gain control behavior, multipath, or platform-specific artifacts). Consequently, data-driven learned methods ranging from classical machine learning to deep neural networks (DNNs) have become prominent for GNSS RFI detection and classification \cite{chen2022gnss, mehr2025deep, van2024optimal, mehr2024dual}. However, many high-accuracy learned methods incur compute and memory footprints that are impractical for embedded receivers or compact adaptive antenna platforms (e.g., CRPA) that must operate within strict real-time and low-resource budgets.

Recent works demonstrated that \emph{expert features} engineered from receiver signal models and compact auxiliary algorithms can improve robustness against harsh conditions and boost accuracy for compact models without requiring massive parameter sets. These features include variations of the Fourier transform (particularly spectrogram images \cite{mehr2025deep}), wavelet transform, Wigner-Ville time-frequency distributions, and other statistical measures on the received signal and its latent representations \cite{chen2022gnss}. Most of the transform-based features provide high classification accuracy thanks to their sparsifying nature, but they also incur a high computational load on the deployment platform (lower than typical DNNs, but still significant). This motivates research into more compact transform and feature extraction computations that would still allow for higher classification accuracy without the significant computational load. Such lightweight expert features can then be used at the input space of the recently popularized online, few-shot, federated and weakly-supervised learning paradigms that offer paths to maintain accuracy across scenarios while reducing labeling burden and privacy risks \cite{wu2023_federated, heublein2024_pseudolabeling, heublein2024_mlrobustness, heublein2025_unsupervised, ott_heublein_icl, wu2025federated}. 

This paper proposes a new feature for lightweight pre-correlation GNSS interference classification. Our central idea is to augment conventional features with statistical latent variables extracted from the frequency estimations of a \emph{virtual adaptive notch filter} (ANF). The regular ANF is a one-pole complex IIR notch with normalized least mean squares (NLMS) updates of the tracking frequency \cite{borio_2006_1polefilter}, designed to invert the narrowband interferer signal. By a virtual ANF, we designate an ANF running in “estimation-only” mode on the raw complex baseband, i.e., the filtering output is not used. The ANF’s NLMS-based frequency tracker produces a proxy for the interferer’s instantaneous frequency $f_i[n]$ and related stationarity statistics (e.g., narrowband vs. broadband signals). We show that these cues are highly informative for both narrowband (incl. chirps) interferers as well broadband ones, and are inexpensive to compute, making them suitable for embedded deployment. While prior works have mostly relied on spectra or time–frequency fingerprints for this, our results show that ANF-derived dynamics can also improve class separability without prohibitively increasing computational cost.

We study the above within a design space constrained by tight compute and memory limits typical of pre-correlation adaptive antenna or receiver modules. Specifically, we target per-frame inference under small memory and compute budgets, and predictable fixed-point performance on microcontrollers and FPGAs. The goal is to preserve accuracy comparable to much larger models while meeting real-time constraints.

Our contributions can be summarized as follows:
\begin{itemize}
    \item We propose a set of lightweight ANF-derived features that complement spectral descriptors and show that they improve classification accuracy for both narrowband and broadband interferers, compared to existing baselines.
    \item We benchmark compact learned algorithms such as shallow gradient-boosted decision trees to demonstrate favorable accuracy–compute-memory trade-offs relative to heavier algorithms used in prior work.
    \item We quantify resource use on a compact adaptive antenna platform (EDGE Microwave HEDGE8008), showing that the solution is suitable for embedded receivers.
\end{itemize}

The rest of the paper is structured as follows: Section II provides the signal model for GNSS interferers and expert-feature based interference classifiers. Section III presents the proposed lightweight virtual ANF-based feature set. Section IV demonstrates the advantages of using the proposed feature on lightweight classifier over a broad suite of simulated and recorded RFI (incl. publicly available datasets) as well as analyzing resource utilization. Section V concludes the paper with a discussion of the results and future work.

\section{GNSS Interference Classification}
\label{sec:classification}

This section formalizes the received-signal model and summarizes feature-extraction and classification approaches that can be used for GNSS interferer classification.

\subsection{Signal Model}
\label{subsec:signal-model}

The discrete-time complex baseband signal observed at the receiver input is expressed as
\begin{equation}
    r[n] = s_{\text{GNSS}}[n] + i[n] + \eta[n],
\end{equation}
where $s_{\text{GNSS}}[n]$ represents the aggregated satellite signals, $\eta[n]$ is additive white Gaussian noise, and $i[n]$ denotes the interference component to be identified. We only consider a single interferer being present at a given time in this article for simplicity. Classification is done by processing $r[n]$ or features extracted from it. $i[n]$ can be one of many types of interferers, including narrowband, broadband and pulsed signals which can be stationary (or not) in their controlling parameters. While there are many different classifications formulated for the general class of interferers \cite{chen2022gnss}, to be able to comprehensively evaluate classifier performance, we assign any interferer to one of the following families in this article:
\begin{equation*}
    \{\texttt{Quiet},~~ \texttt{CW Static}, ~~ \texttt{CW Pulsed}, ~~ \texttt{Chirp Sweep}, ~~ \texttt{Broadband Pulse}, ~~ \texttt{Broadband Noise}, ~~ \texttt{NB Modulated}\},
\end{equation*}
where each family represents a sub-group of similar signal structures and effects on PVT performance. When comparing across datasets, we merge and map other class labels to these families to maintain a unified taxonomy, under the assumption that the results obtained with this mapping should apply to any other narrower class mapping too.

All algorithms considered in this article operate on pre-correlation baseband segments $\mathcal{F}_m$ of $N$ complex samples, using a hop size $H<N$ in between the batches of $N$ complex samples. The window size $N$ is a common parameter across all feature extractors to maintain reliable estimates, carefully chosen to balance time-frequency resolution and latency.

\subsection{Feature Extraction}
\label{subsec:features}

Feature design aims to capture discriminative temporal–spectral patterns in $\mathcal{F}_m$ between classes while remaining computationally tractable for embedded real-time implementation. Following \cite{chen2022gnss} and related studies, we organize features into two groups:

\textbf{Transform Dynamics.}
Power spectral density (PSD) features summarize the energy distribution of $r[n]$ over the analysis window. Typical descriptors include total in-band energy, spectral centroid, bandwidth or spread, roll-off frequency, and spectral flatness. These quantities are extracted from the magnitude spectrum or a short-time Fourier transform (STFT) and normalized by frame energy to provide invariance to absolute signal power. More contrived transforms such as the Wigner-Ville \cite{sun2021_wignervillehough}, Karhunen-Loeve \cite{szumski2013karhunen}, wavelet \cite{sun2021_wavelet} or fractional Fourier exist, with varying amounts of computational loads.

\textbf{Temporal and Statistical Dynamics.}
Frame-to-frame variations of the spectral features, e.g., changes in centroid, bandwidth, or total power, capture nonstationary behavior that differentiates chirps, hops, and pulsed sources from static continuous-wave (CW) tones. Additional low-cost statistics such as skewness, kurtosis, and envelope variance are computed to characterize modulation depth and intermittency.

Due to the high success rate of DNN architectures built for 2D (image) processing, most modern DNN-based methods operate on 2D transform-domain features; the vast majority being spectral features from an STFT. However, expert-feature-based classification shows just as much potential, and with well-designed features, it's more efficient than spectrogram-based inference with DNNs since part of the class representation gets baked into the architecture itself with the expert feature.

\subsection{Classification Algorithms}
\label{subsec:classifiers}

The extracted feature vectors are processed by lightweight learned models optimized for small compute and memory footprint. We consider two algorithm families that are not exhaustive, but highly representative of the useful solutions in this space: 

\begin{itemize}[leftmargin=1em]
    \item \textbf{Baseline Heuristics:} Simple threshold-based or rule-based detectors serve as computational baselines for legacy receivers.
    \item \textbf{Decision-Trees:} Shallow gradient-boosted classifiers offer interpretable decision boundaries and efficient implementations.
\end{itemize}

DNNs are not considered in this study due to the extremely larger resource usage demanded for high accuracy. Expert features can enable much smaller classifiers with decision trees or heuristics-based classifiers since the model is simplified by the algorithm embedded in the expert feature extractor, just like an expert on a topic explaining it to an audience in simpler terms. In this paper we exclusively study the gradient-boosted decision tree (GBDT) algorithms due to their high accuracy and practicality, and show that our proposed expert feature facilitates higher accuracy with lower complexity.

\section{Adaptive Notch Filter Estimation as an Input Feature}

The adaptive notch filter (ANF) is a simple mechanism to estimate and track the instantaneous frequency of a narrowband RFI component. In this work, it is used in a \emph{virtual} estimation-only mode to extract latent feature (instantaneous frequency) characterizing the received signal rather than to perform interference suppression.

We use the canonical complex one-pole ANF with NLMS adaptation, defined as follows \cite{borio_2006_1polefilter}. Let $x[n]\!=\!r[n]$ be the complex baseband input, and let $x_r[n]$ denote the autoregressive (AR) state of the notch (latent variable) with

\begin{equation}
x_r[n] \;=\; x[n] + k_a\,z_0[n{-}1]\;x_r[n{-}1], 
\qquad 
z_0[n]\triangleq e^{j\theta_z[n]},\ \ k_a\in(0,1).
\end{equation}

The notch output is 
\begin{equation}
x_f[n] \;=\; x_r[n] - z_0[n{-}1]\;x_r[n{-}1],
\end{equation}

and we use the instantaneous moving average (MA) + AR output power as the minimization objective,

\begin{equation}
J[n] \;\triangleq\; |x_f[n]|^2.
\end{equation}

Parameterizing with the notch angle $\theta_z$ (and $f_z=\theta_z F_s/(2\pi)$), the gradient of $J[n]$ with respect to $\theta_z$ is

\begin{equation}
g[n] \;\triangleq\; \frac{\partial J[n]}{\partial \theta_z}
\;=\; 2\,\Re\!\left\{\frac{\partial x_f[n]}{\partial \theta_z}\,x_f^*[n]\right\}
\;=\; 2\,\Re\!\Big\{-j\,z_0[n{-}1]\;x_r[n{-}1]\;x_f^*[n]\Big\}.
\end{equation}

The angle update uses the normalized LMS step over a single sample (i.e., stochastic, not batched) gradient

\begin{equation}
\theta_z[n{+}1] \;=\; \theta_z[n] \;-\; \mu[n]\;g[n],
\qquad
\mu[n] \;=\; \frac{\delta}{\widehat{E}\{|x_r[n]|^2\} + \epsilon_\mu},
\end{equation}

with small nominal step $\delta$ and regularizer $\epsilon_\mu>0$. The recursion is run per analysis window $\mathcal{F}_m$ of $N$ samples, using fixed-point friendly arithmetic and a fresh state.

\textbf{Per-window ANF feature vector.} Let $z_0[n]=e^{j\theta_z[n]}$ be the per-sample pole estimate over $\mathcal{F}_m$, and define the unwrapped angle sequence $\phi[n]$ by cumulatively canceling $2\pi$ discontinuities:
\[
\phi[n] \;=\; \phi[n{-}1] + \mathrm{wrap}_{[-\pi,\pi)}\!\big(\theta_z[n]-\theta_z[n{-}1]\big),
\quad \phi[n_0]=\theta_z[n_0].
\]
We form the following statistics (all computed over the window indices in $\mathcal{F}_m$; for brevity we suppress the subscript $m$):

\begin{align}
E \;&=\; \sum_{n\in\mathcal{F}_m} |x[n]|^2 
&& \text{(input energy, power proxy)},\\[3pt]
\bar{\phi} \;&=\; \frac{1}{N}\sum_{n\in\mathcal{F}_m\setminus\{n_{\max}\}}\phi[n], 
\quad 
\sigma_\phi^2 \;=\; \frac{1}{N}\sum_{n}(\phi[n]-\bar{\phi})^2 
&& \text{(mean and variance of angle)},\\[3pt]
\text{Slope} \;&=\; \frac{\sum_{k} (k-\bar{k})\,(\phi[k]-\bar{\phi})}{\sum_{k} (k-\bar{k})^2},
\quad \bar{k}=\tfrac{1}{N}\sum_{k} k
&& \text{(least-squares trend)},\\[3pt]
\phi_c[k] \;&=\; \frac{\phi[k]-\bar{\phi}}{\max(\mathrm{std}(\phi[k]-\bar{\phi}),\,1)}
\quad\Rightarrow\quad
D_1 \;=\; \mathrm{var}\!\big(\Delta \phi_c\big),\;\;
D_2 \;=\; \mathrm{var}\!\big(\Delta^2 \phi_c\big)
&& \text{(1st/2nd diff variances)},\\[3pt]
D_{1,\mathrm{med}} \;&=\; \mathrm{var}\!\big(\Delta\,\mathrm{medfilt}_{L}\{\phi\}\big)
&& \text{(median-filtered diff variance)}.
\end{align}

Here $k$ indexes the $N$ samples used for linear fit, $\Delta$ and $\Delta^2$ denote the first and second forward differences across $k$, and $\mathrm{medfilt}_{L}\{\cdot\}$ is an odd-length median filter applied to $\phi$ before differencing. Normalizing $\phi$ to $\phi_c$ (division by its standard deviation, lower-bounded by 1) improves numerical stability and reduces sensitivity to absolute scaling.

Finally, the ANF feature vector for window $\mathcal{F}_m$ is
\begin{equation}
x^{(\text{ANF})}_m 
\;=\;
\big[\,E,~ \bar{\phi},~ \sigma_\phi^2,~ \text{Slope},~ D_1,~ D_2,~ D_{1,\mathrm{med}}\,\big]^{\top}.
\end{equation}
Intuitively, stationary narrowband (\texttt{CW Static}) yields small $\sigma_\phi^2$, $D_1$, $D_2$ with $\text{Slope}\!\approx\!0$; pulsed CW increases $D_1$ while preserving a stable mean; chirp-like sweeps produce a nonzero \emph{Slope} and elevated $D_1/D_2$; hopping or strongly nonstationary behavior inflates all difference-based variances and suppresses temporal correlation; broadband processes lack a coherent pole trajectory, driving $\phi$ to noise-like fluctuations with large $D_1$ and $D_{1,\mathrm{med}}$. Because the recursion uses only simple complex arithmetic and one normalization per sample, these descriptors add negligible memory and compute cost, while providing discriminative dynamics that complement transform-based features under strict embedded constraints.

\section{Performance Benchmark}

In this section we first analyze the expert signal generated by our proposed virtual ANF, over which we extract the statistical features. Visual analyses over various measured RFI events shows how the ANF frequency tracking parameter is a good class separator. Next, we introduce the three datasets that we use for training and evaluation of classification performance in this study: a simulated dataset covering a wide range of jammer types, one dataset of various recorded RFI events from the \emph{EDGE Microwave} test harness, and one recorded short-snapshot dataset from \cite{heublein2024_mlrobustness}. Using these datasets, we analyze the performance of lightweight GBDT classifiers that combine the proposed virtual ANF features with well-known STFT features for high accuracy. Finally, we present an ablation study that measures the sensitivities of the virtual ANF based method to changes in its hyperparameters $k_a$, $\mu$, and combine that with an analysis of the STFT hyperparameters of transform and hop size.

\subsection{Analysis of the Proposed Virtual ANF-based Expert Features}
\label{subsec:anf-visual}

\begin{figure}
  \centering
  \includegraphics[width=\textwidth]{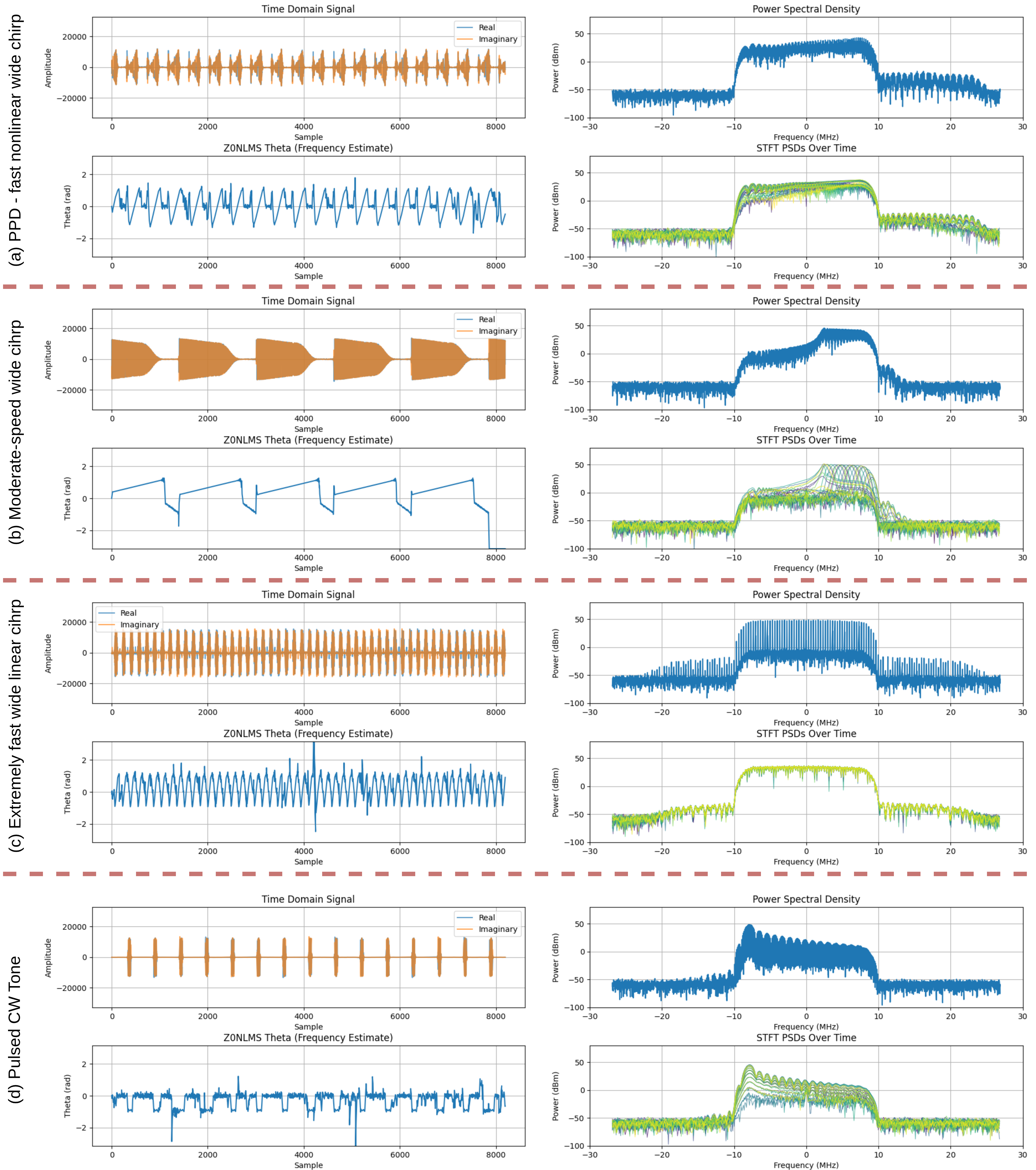}
  \caption{Proposed virtual ANF features plotted against STFT features as well as full-window time and PSD for various jammer types.}
  \label{fig:features1}
\end{figure}

\begin{figure}
  \centering
  \includegraphics[width=\textwidth]{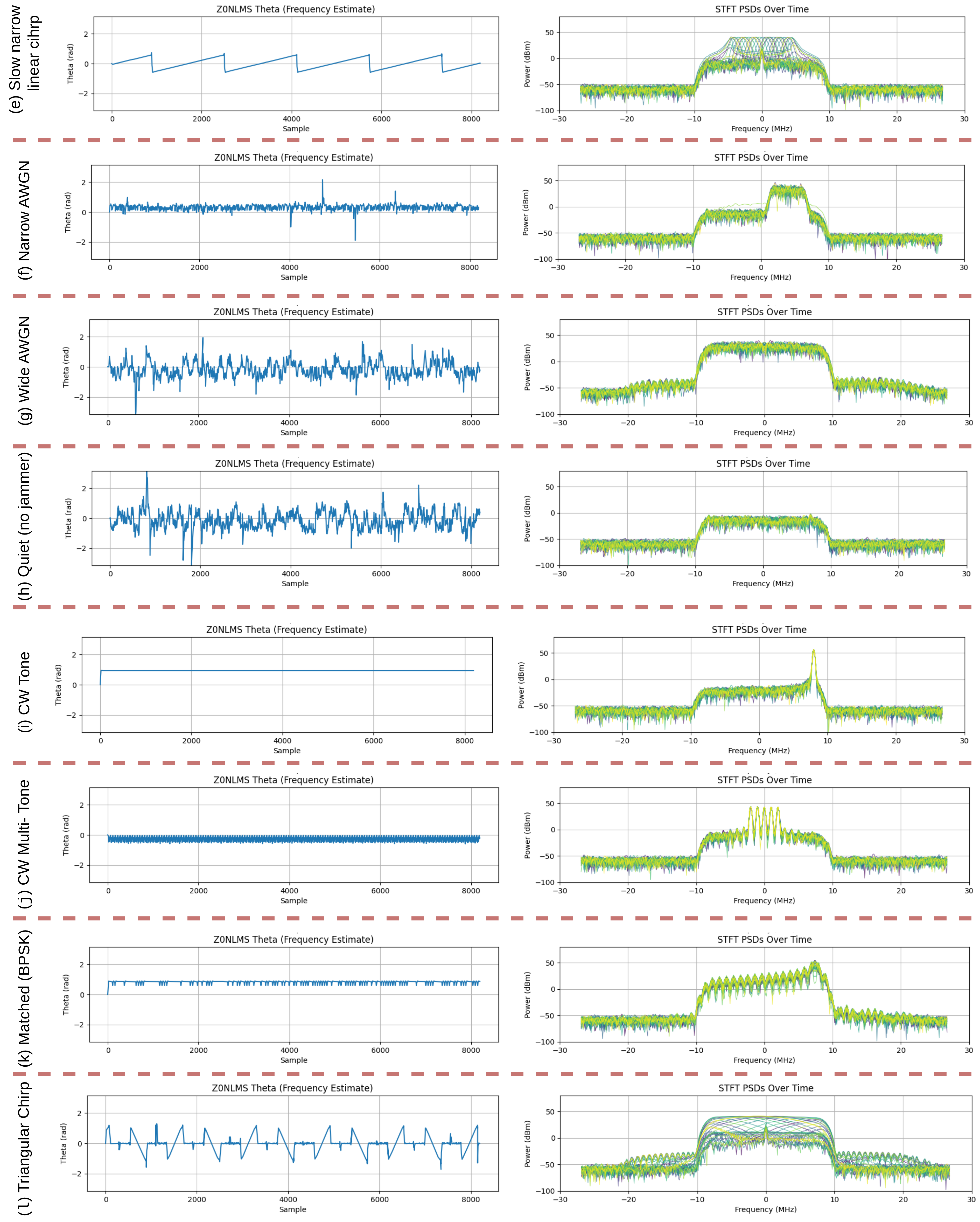}
  \caption{Proposed virtual ANF features plotted against STFT features for various jammer types.}
  \label{fig:features2}
\end{figure}

Figs.~\ref{fig:features1}--\ref{fig:features2} show, (i) the raw time series and (ii)  the full-window PSD where available, as well as (iii) the per-sample ANF pole angle $\theta_z[n]$ and (iv) the classical STFT bin features, for a representative set of interferers measured over the \emph{EDGE Microwave} test harness. The STFT has a high-overlap configuration with a transform length of 128 and a hop length of 16 to catch fast jammers with high resolution. The ANF uses $k_a=0.8~,~\mu=0.04$. We next discuss and compare the behavior of the ANF and STFT features for each jammer type.

Panels (a)-(c) and (e), (l) show various chirps. In each chirp, the $\theta_z[n]$ exhibits a clean ramps that can be captured with a slope measurement and $D_1$/$D_2$ measurements between the reset points (where the chirp falls out of band, e.g., in (a) and (b)). The separation of these clean ramps from other signals is more apparent than that of the STFT bins, which lose their distinguishing ability with higher chirp speeds like in (c). Clean and moderate-speed chirps like in (e) can be distinguished by both STFT and the proposed ANF with high accuracy.

Panel (i) shows a single tone in which $\theta_z[n]$ locks and remains flat with very small variance ($\sigma_\phi^2\!\approx\!0$) and $D_1,D_2\!\approx\!0$. Due to the sparsity in time and frequency, this case is easily distinguishable via both ANF and STFT features. In (d), the tone is periodically switched on and off, creating a pulsed interferer. While the STFT struggles to recognize this behavior due to the full-band coverage of the pulsing action, the ANF catches this by intermittently switching between noise behavior and tone-locking behavior (around $\theta_z \approx -1$), revealing the duty cycle. This behavior is clearly advantageous since it separates the pulsed tone from the BPSK matched-jammer in (k), which is not easy to do over the STFT features. Multi-tone CW in (j) presents several closely spaced lines in STFT, but ANF fails to track multiple tones at the same time due to its narrowband modeling. The GPS-like narrowband modulated emission (panel (k), “Matched (BPSK)”) creates a comb-like PSD; the ANF tracks the fundamental and jitters at phase switching points, creating a statistical signature for recognizing the BPSK jammer, and the STFT features resemble the pulsed tone, showcasing the advantage of the ANF-based approach here. 

Panels (f)–(h) and (g) portray noise-like cases, quiet (no jammer), narrow AWGN and wide AWGN. In both AWGN cases, the STFT lacks coherent tones, and $\theta_z[n]$ behaves as a bounded random walk within the AWGN bandwidth coverage, with near-zero mean correlation across samples, leading to large $D_1$ and reduced $\rho_{f}$. The quiet case in (h) naturally looks exactly the same as the wide AWGN case since it's just an elevated noise floor, however it's easily recognizable via an in-band energy measurement, owing to the fact that the noise floor can easily be calibrated for the quite case.

In summary, the ANF and the STFT feature sets clearly complement each other, in support of the idea of using ANF and STFT together for classification algorithms since they cover each others' weak-performing scenarios. We next benchmark lightweight classifiers across these feature sets to support this claim.

\subsection{Generalization Capability}

We demonstrate the utility of the proposed virtual ANF features over feature extractor - classifier combinations. To maintain the low resource utilization goal, we only benchmark a few different sizes of GBDTs trained with the XGBoost library \cite{chen2016xgboost}. We use a training split from one dataset (the \emph{EDGE Microwave} synthetic RFI dataset) to train/tune the models, and the rest for testing generalization performance. 

We utilize three datasets: (i) \emph{EDGE Microwave} synthetic RFI generated with a matched RF+digital front-end simulator that produces complex baseband snapshots over a wide span of jammer parameterizations ($N=8192,~F_s=60$MHz); (ii) \emph{EDGE Microwave} test harness recordings collected with the actual front-end ($N=8192,~F_s=60$MHz), including several commercial PPD captures with fast nonlinear sweeps; (iii) the DARCY Spectrum Controlled Low-Frequency dataset recorded at Fraunhofer IIS L.I.N.K. ($N=1024,~F_s=102.4$MHz), from which we select a subset of labeled examples that fit our setting and map to the jammer family taxonomy considered in this paper. 

The \emph{EDGE Microwave} synthetic RFI dataset contains 200.000 samples in total, where $\approx15\%$ of those samples are from "quiet" scenarios containing only the useful GNSS signals and noise, and the rest are various single-interferer scenarios, constituting a balanced coverage among the different classes defined in Section II. The second hardware-based \emph{EDGE Microwave} dataset contains 259 recordings of RFI events recorded in an isolated lab environment, where similarly 10 of those samples are for quiet cases and the rest covers a wide variety of jammer types, including commercial PPDs. Information on the DARCY dataset is provided in \cite{heublein2024_mlrobustness}; to focus on the primary proposition of the paper, we filtered the DARCY dataset according to the classes we consider in this paper, ending up with $\approx$ 3500 recordings of 10$\mu$s long RFI event snapshots. Training on the large synthetic RFI dataset and testing on the other two allows us to demonstrate the generalization capability of even a lightweight learner when a good simulation of the actual hardware is used.

Table~\ref{tab:gen-overall} summarizes the overall family-level accuracy for each feature configuration. The ANF-only model has worse performance than FFT, but combining ANF and FFT features improves the accuracy, supporting the hypothesis that the two feature sets complement each other. The trend also applies to the other two datasets, strengthening the justification. Also note that the DARCY dataset is considerably harder than the other two datasets, with an 8x shorter window size, a high prevalence of slow-moving nonstationary jammers to compound the effect of the short window, and lower jammer power to noise ratios, hence the lower accuracy.

\begin{table}[h]
\centering
\caption{Overall family-level accuracy (\%) of the z2 classifier with different feature sets. The model is trained once on the synthetic \emph{EDGE Microwave} dataset and evaluated on three datasets where the synthetic dataset evaluation is on the validation split.}
\label{tab:gen-overall}
\begin{tabular}{lccc}
\hline
Feature set & EDGE - Synthetic & EDGE - recorded & DARCY \\
\hline
ANF only   & 61.32 & 67.95 & 41.28 \\
FFT only   & 75.52 & 77.61 & 57.29 \\
ANF + FFT  & \textbf{79.46} & \textbf{79.92} & \textbf{59.12} \\
\hline
\end{tabular}
\end{table}

Next, we further analyze the performance gain from the ANF-based features. The results are shown on Table~\ref{tab:gen-perclass}, which demonstrates the per-family accuracy of classifiers based on each feature set, on a new instance of the synthetic dataset (different random seed). In all cases we observe that ANF+FFT either matches the better of ANF-only and FFT-only or provides additional gains, supporting the use of ANF features as a complementary low-cost descriptor rather than a stand-alone replacement for spectral features.

\begin{table}[h]
\centering
\caption{Per-family accuracy (\%) on the validation split of the synthetic \emph{EDGE Microwave} dataset for the z2 classifier trained with ANF-only, FFT-only, and combined ANF+FFT features.}
\label{tab:gen-perclass}
\begin{tabular}{lccc}
\hline
Family            & ANF only & FFT only & ANF + FFT \\
\hline
\texttt{Quiet}           & 99.65 & 99.63 & 99.74 \\
\texttt{CW Static}       & 47.83 & 61.77 & 66.49 \\
\texttt{CW Pulsed}       & 41.23 & 62.86 & 74.13 \\
\texttt{Chirp Sweep}     & 65.09 & 70.01 & 74.89 \\
\texttt{Broadband Pulse} & 83.25 & 79.91 & 83.02 \\
\texttt{Broadband Noise} & 35.46 & 59.55 & 63.29 \\
\texttt{NB Modulated}    & 56.62 & 92.26 & 92.19 \\
\hline
\end{tabular}
\end{table}

Overall, these results indicate that the proposed ANF-derived statistics generalize well when combined with standard FFT-based features. They provide additional discriminative information on several jammer families and yield consistent improvements in accuracy across both synthetic and recorded datasets, while remaining compatible with a low-complexity GBDT classifier suitable for embedded deployment.


\subsection{Complexity–Accuracy Tradeoff}
\begin{figure}[h]
\centering
\includegraphics[width=0.99\linewidth]{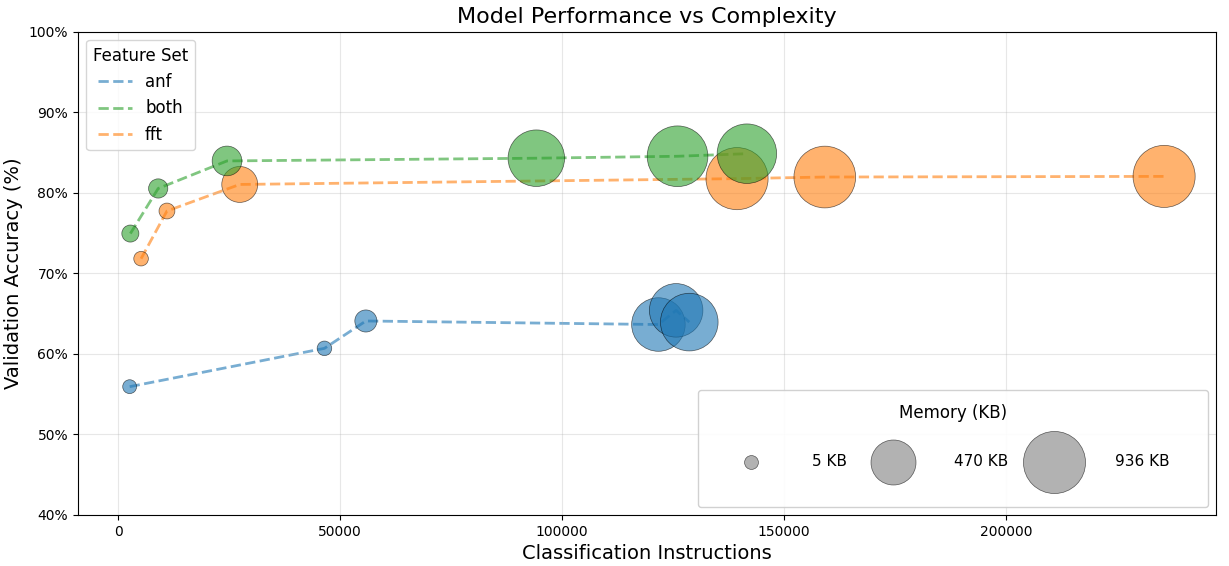}
\caption{Accuracy versus model complexity for the ANF-only, FFT-only, and ANF+FFT feature sets. The horizontal axis represents compute cost, bubble size encodes model memory, and the vertical axis shows validation accuracy on the 20k synthetic subset. Trend lines for each feature set highlight the differing saturation behaviors.}
\label{fig:tradeoff}
\end{figure}

The compactness of a classifier can be gauged by how high an accuracy it can attain with fewer parameters and reduced compute, and better expert features enable algorithms that are more compact. In this regard, for the three feature input options (FFT-only, ANF-only, and ANF+FFT), we train different-sized classifiers to observe the trend in accuracy versus model complexity. Performance is measured using the in-set validation accuracy on a 20k-sample synthetic subset, while model size is quantified in terms of estimated memory footprint and number of operations required for inference. Figure~\ref{fig:tradeoff} summarizes these experiments using a bubble plot, where the horizontal axis corresponds to compute cost, bubble size encodes memory usage, and the vertical axis shows classifier accuracy.

Across all measured complexity levels, the relative ranking of the feature sets remains stable. ANF-only models consistently produce the lowest accuracy and also exhibit an early saturation point: even substantial increases in model size yield diminishing or negligible accuracy improvements. FFT-only models achieve substantially higher accuracy at every complexity level, and their accuracy increases smoothly with model size before plateauing at moderate-to-high complexity ranges. The combination of ANF and FFT features (ANF+FFT), however, provides the strongest performance throughout the entire sweep. Even when the classifier is extremely compact, ANF+FFT outperforms both individual feature sets, and this advantage persists as the models grow larger. The accuracy curve for ANF+FFT rises more steeply at low complexity, reaches higher absolute performance, and saturates later than either ANF or FFT alone.

Another notable trend is that the marginal benefit of enlarging the classifier is strongly dependent on the feature set. For ANF-only, additional depth or boosting rounds quickly provide diminishing returns, indicating that the representational limitations of the features dominate, and this is seen by the multiple trials of models getting stacked together near the same point due to an early stopping of the training (generalization performance does not increase further with size, meaning it's a fundamental limit of the input space). For FFT-only models, increased capacity helps capture more spectral structure, but improvements become modest beyond moderate complexity. In contrast, ANF+FFT maintains meaningful gains longer into the high-complexity regime, indicating that the two feature sources contribute complementary information that the classifier can leverage when afforded additional capacity.

From a deployment perspective, particularly in resource-limited embedded contexts, these results highlight the advantage of using hybrid features. For a fixed compute or memory budget, ANF+FFT enables significantly higher accuracy than either input set alone. Conversely, if a target accuracy threshold is predetermined, ANF+FFT typically reaches that level at substantially lower model sizes, enabling reductions in compute cycles, model storage, and power consumption.



\subsection{Ablation Study - Sensitivity Against Hyperparameters}

The hyperparameters of the feature extraction methods we've discussed so far are the $k_a$ and $\mu$ of the ANF, and the transform and hop sizes of the FFT. The sensitivity of the whole classification accuracy to these hyperparameters is an important performance attribute since these parameters aren't trained at development time, thus, they require calibration. While naturally $N$ is also a hyperparameter, its choice is trivial since the complexity of the feature extraction directly scales with it (linearly with ANF, and $N\log N$ with FFT), and the window needs to be long enough to capture the characteristics of the jammer when it is non-stationary. 

\begin{figure}[h]
\centering
\includegraphics[width=0.99\linewidth]{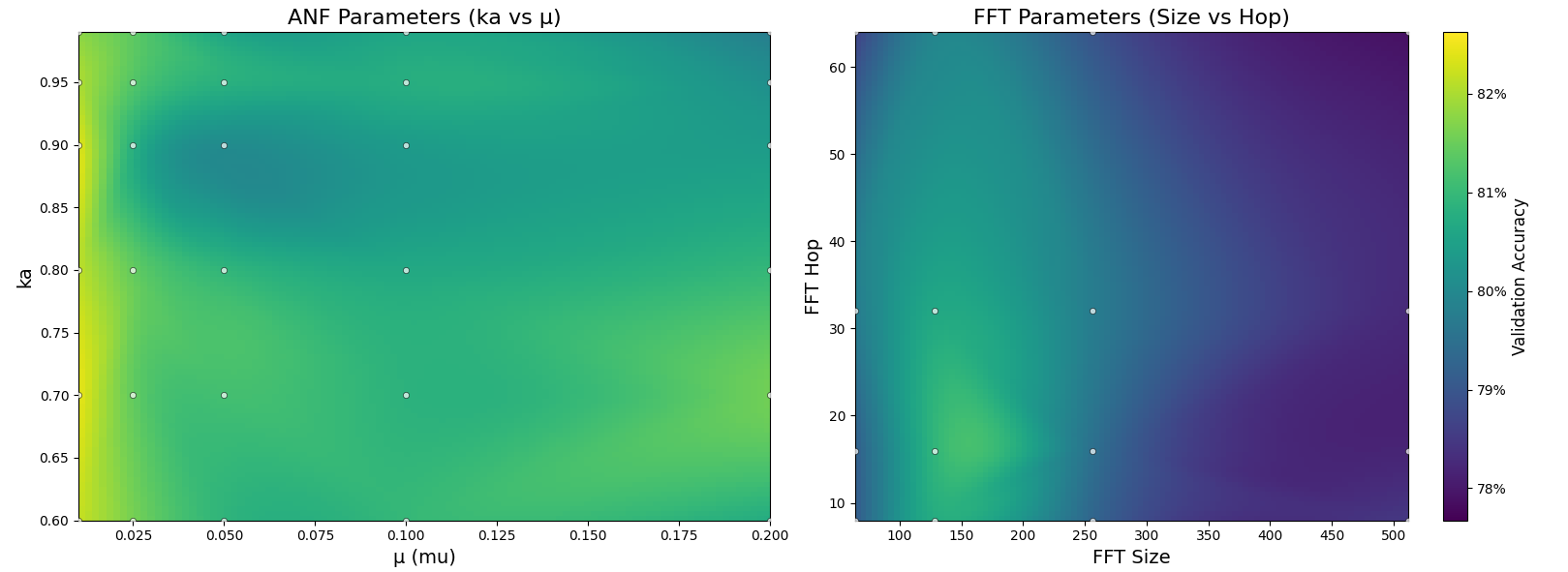}
\caption{Accuracy versus ANF and FFT hyperparameters. The effects of the hyperparameter variations are minor since they move the accuracy at around $\pm$2\% maximum.}
\label{fig:hyperparams}
\end{figure}

Figure \ref{fig:hyperparams} demonstrates the effects of a sweep of these four hyperparameters on the validation split of the 20k-sample synthetic RFI dataset. The results show that the effects of the hyperparameters are minor, but they are still effective. The best performance for this dataset, which covers a wide range of jammer control parameters from very fast chirps to very slow pulsed signals and random noise, turns out to be at a transform size of 128, hop size of 16, $k_a=0.8$ and $\mu=0.015$. 

\section{Conclusion}

This paper considered real-time pre-correlation GNSS interference classification under tight compute and memory constraints. We introduced a set of lightweight expert features based on the instantaneous frequency tracking behavior of a virtual adaptive notch filter (ANF), and combined them with conventional FFT-based spectral descriptors. The ANF is operated purely in estimation mode and provides a small number of statistics that capture narrowband versus broadband behavior, chirp-like dynamics, and pulsing, while remaining inexpensive to compute and amenable to fixed-point implementation.

Using synthetic, hardware-recorded, and publicly available datasets, we showed that ANF-only features are weaker than FFT-only features in isolation, but that the combined ANF+FFT feature set consistently improves classification accuracy across all datasets and jammer families. The gains hold for compact gradient-boosted decision tree (GBDT) models and persist across a range of model sizes, indicating that the ANF-derived statistics provide complementary information to FFT features. A complexity–accuracy analysis further showed that ANF+FFT reaches a given accuracy at lower model complexity than FFT-only, and that the overall performance is relatively insensitive to moderate changes in ANF and FFT hyperparameters.

Overall, the results indicate that virtual ANF-based features are a practical addition to lightweight pre-correlation classifiers for embedded GNSS receivers and compact CRPA platforms. Future work will extend the approach to multi-interferer scenarios, tighter integration with adaptive mitigation chains in the receiver, and online or weakly-supervised learning schemes that can adapt the classifier to changing interference environments while preserving low resource usage.

\section*{acknowledgements}

\noindent The authors thank the \textit{EDGE Microwave} team for their contributions to the development and testing of the system. In particular, we are grateful to Yusuf \c{S}ahin and Abdulkadir Uzun for their support with jammer simulations, test harness operation, data collection, and many helpful discussions.

\bibliographystyle{apalike}
\bibliography{references}

@techreport{eu_sota_rfi_2023,
  title       = {STATE OF THE ART OF GNSS RFI DETECTION, MITIGATION AND LOCALIZATION TECHNIQUES (AIRING Project Report REP-03)},
  author      = {{European Commission}; GMV},
  institution = {European Commission},
  address     = {Brussels},
  year        = {2021},
  month       = {September},
  note        = {AIRING-GMV-D2040-REP-03 V1.1; cited in AIRING Final Report (2023).}
}

@inproceedings{vandermerwe_ppd_eval_2018,
  title={Evaluation of mitigation methods against COTS PPDs},
  author={van der Merwe, J Rossouw and R{\"u}gamer, Alexander and Garzia, Fabio and Felber, Wolfgang and Wendel, Jan},
  booktitle={2018 IEEE/ION Position, Location and Navigation Symposium (PLANS)},
  pages={920--930},
  year={2018},
  organization={IEEE}
}

@article{borio_benchmark_2020,
  title={GNSS interference mitigation: A measurement and position domain assessment},
  author={Borio, Daniele and Gioia, Ciro},
  journal={NAVIGATION: Journal of the Institute of Navigation},
  volume={68},
  number={1},
  pages={93--114},
  year={2021},
  publisher={Institute of Navigation}
}

@article{bek2015_cairo_postcorr,
  title={Classification and mathematical expression of different interference signals on a GPS receiver},
  author={Bek, Mohammad K and Shaheen, Ehab M and Elgamel, Sherif A},
  journal={Navigation: Journal of The Institute of Navigation},
  volume={62},
  number={1},
  pages={23--37},
  year={2015},
  publisher={Wiley Online Library}
}

@article{kim2012gnss,
  title={A GNSS interference detection method based on multiple ground stations},
  author={Kim, Sun Young and Kang, Chang Ho and Yang, Jeong Hwan and Park, Chan Gook and Joo, Jung Min and Heo, Moon Beom},
  journal={Journal of Positioning, Navigation, and Timing},
  volume={1},
  number={1},
  pages={15--21},
  year={2012},
  publisher={The Institute of Positioning, Navigation, and Timing}
}

@article{borio_2014_magazine,
  title={Fast and flexible tracking and mitigating a jamming signal with an adaptive notch filter},
  author={Borio, Daniele and O'DRISCOLL, Cillian and FORTUNY, GUASCH Joaquim and others},
  journal={Inside GNSS},
  year={2014}
}

@article{chen2022gnss,
  title={GNSS interference type recognition with fingerprint spectrum DNN method},
  author={Chen, Xin and He, Di and Yan, Xinyu and Yu, Wenxian and Truong, Trieu-Kien},
  journal={IEEE Transactions on Aerospace and Electronic Systems},
  volume={58},
  number={5},
  pages={4745--4760},
  year={2022},
  publisher={IEEE}
}

@inproceedings{wu2023_federated,
  title={Jammer classification with federated learning},
  author={Wu, Peng and Calatrava, Helena and Imbiriba, Tales and Closas, Pau},
  booktitle={2023 IEEE/ION Position, Location and Navigation Symposium (PLANS)},
  pages={228--234},
  year={2023},
  organization={IEEE}
}

@article{heublein2024_pseudolabeling,
  title={Achieving Generalization in Orchestrating GNSS Interference Monitoring Stations Through Pseudo-Labeling},
  author={Heublein, Lucas and Feigl, Tobias and R{\"u}gamer, Alexander and Ott, Felix},
  journal={arXiv preprint arXiv:2410.14686},
  year={2024}
}

@article{heublein2024_mlrobustness,
  title={Evaluating ML robustness in GNSS interference classification, characterization \& localization},
  author={Heublein, Lucas and Feigl, Tobias and Nowak, Thorsten and R{\"u}gamer, Alexander and Mutschler, Christopher and Ott, Felix},
  journal={arXiv preprint arXiv:2409.15114},
  year={2024}
}

@inproceedings{borio_2006_1polefilter,
  title={Analysis of the one-pole notch filter for interference mitigation: Wiener solution and loss estimations},
  author={Borio, Daniele and Camoriano, Laura and Mulassano, Paolo},
  booktitle={Proceedings of the 19th International Technical Meeting of the Satellite Division of The Institute of Navigation (ION GNSS 2006)},
  pages={1849--1860},
  year={2006}
}

@article{van2024optimal,
  title={Optimal machine learning and signal processing synergies for low-resource GNSS interference classification},
  author={van der Merwe, Johannes Rossouw and Franco, David Contreras and Feigl, Tobias and R{\"u}gamer, Alexander},
  journal={IEEE Transactions on Aerospace and Electronic Systems},
  volume={60},
  number={3},
  pages={2705--2721},
  year={2024},
  publisher={IEEE}
}

@article{mehr2025deep,
  title={A deep neural network approach for classification of GNSS interference and jamming},
  author={Mehr, Iman Ebrahimi and Dovis, Fabio},
  journal={IEEE Transactions on Aerospace Electronic Systems},
  volume={61},
  number={2},
  pages={1660--1676},
  year={2025}
}

@inproceedings{mehr2024dual,
  title={Dual-stage deep learning approach for efficient interference detection and classification in GNSS},
  author={Mehr, Iman Ebrahimi and Savolainen, Outi and Ruotsalainen, Laura and Dovis, Fabio},
  booktitle={Proceedings of the 37th International Technical Meeting of the Satellite Division of The Institute of Navigation (ION GNSS+ 2024)},
  pages={3336--3347},
  year={2024}
}

@article{heublein2025_unsupervised,
  title={Evaluation of (un-) supervised machine learning methods for GNSS interference classification with real-world data discrepancies},
  author={Heublein, Lucas and Raichur, Nisha L and Feigl, Tobias and Brieger, Tobias and Heuer, Fin and Asbach, Lennart and R{\"u}gamer, Alexander and Ott, Felix},
  journal={arXiv preprint arXiv:2503.23775},
  year={2025}
}

@inproceedings{ott_heublein_icl,
  author = {Felix Ott and Lucas Heublein and Nisha Lakshmana Raichur and Tobias Feigl and Jonathan Hansen and Alexander Rügamer and Christopher Mutschler},
  title = {{Few-Shot Learning with Uncertainty-based Quadruplet Selection for Interference Classification in GNSS Data}},
  booktitle = {\href{https://ieeexplore.ieee.org/document/10578525}{IEEE Intl. Conf. on Localization and GNSS (ICL-GNSS)}},
  month = jun,
  year = {2024},
  address = {Antwerp, Belgium},
  doi = {10.1109/ICL-GNSS60721.2024.10578525}
}

@article{sun2021_wignervillehough,
  title={A novel GNSS interference detection method based on Smoothed Pseudo-Wigner--Hough transform},
  author={Sun, Kewen and Yu, Baoguo and Elhajj, Mireille and Ochieng, Washington Yotto and Zhang, Tengteng and Yang, Jianlei},
  journal={Sensors},
  volume={21},
  number={13},
  pages={4306},
  year={2021},
  publisher={MDPI}
}

@article{sun2021_wavelet,
  title={A new GNSS interference detection method based on rearranged wavelet--hough transform},
  author={Sun, Kewen and Zhang, Tengteng},
  journal={Sensors},
  volume={21},
  number={5},
  pages={1714},
  year={2021},
  publisher={MDPI}
}

@inproceedings{szumski2013karhunen,
  title={The Karhunen-Loeve transform as a future instrument to interference mitigation},
  author={Szumski, A and Eissfeller, B},
  booktitle={Proceedings of the 26th International Technical Meeting of the Satellite Division of The Institute of Navigation (ION GNSS+ 2013)},
  pages={3443--3449},
  year={2013}
}

@inproceedings{chen2016xgboost,
  title={Xgboost: A scalable tree boosting system},
  author={Chen, Tianqi and Guestrin, Carlos},
  booktitle={Proceedings of the 22nd acm sigkdd international conference on knowledge discovery and data mining},
  pages={785--794},
  year={2016}
}

@article{wu2025federated,
  title={Federated learning of jamming classifiers: From global to personalized models},
  author={Wu, Peng and Calatrava, Helena and Imbiriba, Tales and Closas, Pau},
  journal={NAVIGATION: Journal of the Institute of Navigation},
  volume={72},
  number={1},
  year={2025},
  publisher={Institute of Navigation}
}

\end{document}